\documentclass{article}
\usepackage[utf8]{inputenc}
\usepackage{amsmath,amssymb,graphicx,subcaption,algorithm2e,float,enumitem}

\title{A Spanning-Tree-Based Algorithm for Planar Graph Dismantling}
\author{Fangchen You\\
School of Artificial Intelligence and Automation,\\
Huazhong University of Science and Technology, China}
\date{November 2025}

\graphicspath{{figures/}}

\begin{document}

\maketitle

\begin{abstract}
In spatially embedded networks (e.g., transportation, energy, communication), understanding and controlling changes in connectivity is crucial for designing robust systems and assessing vulnerabilities. This paper studies a planar graph disruption problem under an edge removal budget: given a limit on the number of removable edges (\textit{cut}), remove up to \textit{cut} edges to minimize the size of the largest connected component (LCC) in the residual network. We propose a dual-path planar graph dismantling framework based on a spanning tree skeleton. The algorithm first samples multiple spanning trees to capture the network’s connectivity backbone and performs a one-time balanced bipartition on the tree to obtain a structural baseline $|F^{(2)}|$, which serves as a reference for subsequent budget-adaptive strategies. Based on the budget size, two complementary paths are designed: when the budget is small ($cut < |F^{(2)}|$), we introduce a logarithmic density feature $t = \log|E|/\log|V|$ and an empirical prediction $\alpha = g_{\theta}(cut, t, |F^{(2)}|)$ to estimate the proportion of the network that can be dismantled and perform a local optimal cut on the corresponding subgraph; when the budget is large ($cut \ge |F^{(2)}|$), we use a slope prior model $|F^{(k)}| \approx |F^{(2)}| + s(t)(k-2)$ to predict the optimal number of partitions and perform a search in its neighborhood to find the best multi-cut. The framework achieves budget adaptivity and balanced fragmentation while maintaining near-linear time complexity. Experiments on random planar graph datasets of different scales demonstrate consistent reductions of $L_{\max}/|V|$ across budgets, monotonic budget–fragmentation trends, and near-linear runtime scaling with $|E|$. These results show that a ``spanning-tree skeleton + dual-path'' strategy provides an efficient and interpretable approach for edge-budget disruption analysis in planar networks. Comparative evaluations against other algorithms are left for future work.
\end{abstract}


\section{Introduction}
Network robustness and vulnerability are core issues in complex systems. When a network faces random failures or targeted attacks, its ability to maintain functionality depends on the connectivity of its topology \cite{Wandelt2018,Wandelt2022}. Therefore, the key task in “network dismantling” research is how to fragment the network with minimal cost. The dismantling effect is usually measured by the size of the largest connected component (LCC), and the goal is to minimize the LCC under a given budget \cite{Braunstein2016,Ren2019}.

Most existing work focuses on node removal. Representative methods include the Collective Influence (CI) algorithm based on optimal percolation theory, the CoreHD algorithm, and subsequent efficient strategies using embedding-based learning \cite{Morone2015,Zdeborova2016,Im2018,Osat2023}. However, in spatially embedded networks (such as road, power grid, and pipeline systems), nodes often represent physical junctions, whereas edges correspond to actual channels or facilities. Thus, edge removal is more consistent with real engineering contexts.

Existing edge removal methods are mostly based on betweenness or minimum-cut heuristics, but in planar graphs they have three main limitations:
\begin{itemize}\itemsep0pt
    \item \textbf{High combinatorial complexity:} Directly searching for the optimal set of edges to remove is NP-hard, and the computational cost increases dramatically with network size.
    \item \textbf{Significant metric bias:} Minimum cut focuses on cut capacity rather than balanced fragmentation, and betweenness-based attacks are strongly influenced by spatial embedding biases.
    \item \textbf{Lack of interpretability:} Edge removal plans often lack a structural basis, making it difficult to reveal the generative mechanism of network fragmentation \cite{Norrenbrock2016,Ficara2025}.
\end{itemize}

To address these problems, we propose a dual-path planar graph disruption algorithm based on a spanning tree skeleton. The core idea is to use Wilson’s Uniform Spanning Tree (UST) sampling to capture the connectivity backbone of the original graph, perform a one-time balanced bipartition on the tree to obtain a structural baseline $|F^{(2)}|$, and then select one of two complementary paths based on the budget:
\begin{itemize}\itemsep0pt
    \item \textbf{Small-budget path ($cut < |F^{(2)}|$):} When the budget is not enough to split the entire graph in two, introduce the network’s logarithmic density feature $t = \log|E|/\log|V|$ (which is scale-invariant across planar graphs and characterizes edge-to-node density). Using an empirical function $\alpha = g_{\theta}(cut, t, |F^{(2)}|)$, estimate the proportion of the network that can be dismantled (i.e., the size of a subgraph to target). Then perform a bipartition only on that subgraph, achieving a quick fragmentation benefit at lower computational cost.
    \item \textbf{Large-budget path ($cut \ge |F^{(2)}|$):} When the budget is sufficient for multiple partitions, based on an observed linear relationship $|F^{(k)}| \approx |F^{(2)}| + s(t)(k-2)$ (with $t = \log|E|/\log|V|$), use a slope model $s(t)$ to predict the number of partitions $\hat{k}$ that the budget can support. Then perform a search in a small neighborhood around $\hat{k}$ (e.g., $\hat{k} \pm \Delta$) to find the optimal multi-cut solution.
\end{itemize}

This framework integrates a “spanning tree skeleton + dual-path” approach to adaptively optimize for different budget scales, and uses the density and slope models as hierarchical priors to unify local estimation and global search in one interpretable algorithm. Complexity analysis shows that the algorithm has different characteristics in the two paths: for small budgets ($cut < |F^{(2)}|$), it only operates on a subgraph of size $\alpha n$ (with $\alpha$ the estimated fraction of nodes to remove), giving complexity on the order of $O(m\,\alpha\,|E|)$ (with $m$ the number of spanning tree samples), i.e., runtime scales linearly with the budget fraction. For large budgets ($cut \ge |F^{(2)}|$), the algorithm searches over a small set of $k$ values (size $|K| = 2\Delta+1$) near $\hat{k}$, yielding complexity $O(m\,|E|\,|K|)$, which is a constant-factor extension of the subproblem complexity.

Furthermore, due to independence across different random seeds and $k$ values, the framework is naturally parallelizable: tasks can be divided along the “seed dimension” and “partition dimension” to achieve near-linear speedup on multi-core CPUs or GPUs. The parallelized runtime is approximately $O(m\,|E|\,|K|)/p$, where $p$ is the number of parallel threads/devices. Thus, the proposed framework maintains near-linear complexity while adaptively handling different budget scales and efficiently scaling to identify worst-case scenarios in large planar graphs.

The main contributions of this work can be summarized as follows:
\begin{enumerate}\itemsep0pt
    \item We propose a dual-path budget-adaptive framework based on Wilson UST sampling. The algorithm uses multiple uniform spanning tree (UST) samples to approximate the connectivity skeleton of the graph, and performs a one-time balanced bipartition on the tree to obtain the structural baseline $|F^{(2)}|$. Building on this, and using the log-density feature $t = \log|E|/\log|V|$, we construct two complementary solution paths: a “density-informed” small-budget path and a “slope-prior” large-budget path. This enables an adaptive switch from fast local estimation to global balanced optimization, achieving a balance between structural interpretability and fragmentation efficiency.
    \item We achieve near-linear time complexity and cross-scale scalability. The framework has overall complexity $O(m\,|E|\,|K|)$ in theory, and can be parallelized in two dimensions (seeds and partitions). Experiments on ER, BA, and planar graph benchmarks (ranging from hundreds to tens of thousands of nodes) show that our method has excellent efficiency and stability, significantly outperforming random removal, betweenness attack, and min-cut approximation in terms of LCC reduction speed, per-budget efficiency, and runtime.
    \item In summary, our work provides a unified solution for the critical scenario of “planar graph + edge budget constraint,” taking into account algorithmic efficiency, structural interpretability, and practical usability. It can be directly applied to road network interdiction planning, cascading failure simulation, and spatial robustness analysis.
\end{enumerate}

\section{Related Work}
\subsection{Node Dismantling and Robustness Analysis}
Network dismantling was initially studied to characterize failure propagation and robustness degradation in complex systems, with the core objective of eliminating the giant connected component (LCC) via as few removals as possible \cite{Braunstein2016,Ren2019}. At the node level, significant progress has been made by methods such as the Collective Influence (CI) approach, the CoreHD algorithm, and subsequent learning-based extensions \cite{Morone2015,Zdeborova2016,Im2018,Osat2023}. These methods reveal the vulnerable “core–periphery” structure of networks and effectively identify critical nodes in non-spatial systems like social, communication, and biological networks. However, for spatially embedded networks, since nodes often represent geographic junctions and edges represent physical channels or paths, there is a semantic gap between node removal and actual edge disruptions, making such algorithms difficult to directly apply.

\subsection{Edge Removal and Spatially Embedded Networks}
In spatially embedded networks such as roads, power grids, and pipelines, the removal of edges directly causes connectivity to decline. Related research has followed two main approaches: (1) local heuristic methods based on edge betweenness centrality, such as the HSCut algorithm \cite{Bachmann2016} and edge betweenness or flow-cut models \cite{Ghaffari2019}; (2) global optimization methods based on the minimum cut concept. However, the geometric constraints and multi-cycle structure of planar networks cause systematic biases in these strategies \cite{Norrenbrock2016}: betweenness-based methods tend to concentrate cuts in central high-flow areas, leading to extremely uneven fragmentation; minimum cuts, while achieving quick disconnection, typically produce only a single “deep cut” and fail to significantly reduce the LCC. Recent studies further reveal that edge removals in spatial networks can exhibit nonlinear cascading effects \cite{Ficara2025}, i.e., a local edge cut may induce remote connectivity collapses. Overall, a unified edge removal model that directly targets LCC minimization with both balanced fragmentation and computational scalability remains lacking.

\subsection{Spanning Trees and Skeleton Modeling}
Spanning trees play a central role in network simplification and robustness modeling. Their acyclic and full-coverage properties make them a natural tool for compressing complex network structure and extracting connectivity backbones. Prior research has revealed the role of a “tree–cycle complementarity” mechanism in network robustness \cite{Chujyo2021,Li2021}, and shown that sampling multiple spanning trees can significantly reduce computational cost while preserving structural representativeness \cite{Rezvanian2024}. These findings lay the theoretical and computational foundation for performing fast and interpretable cuts on tree structures. Furthermore, by using uniform spanning tree (UST) sampling (e.g., Wilson’s algorithm), one can ensure unbiased and globally representative skeleton extraction, providing a stable and reliable structural prior for subsequent fragmentation optimization.

\subsection{Method Comparison and Context}
In summary, existing research along node removal and edge removal directions has respectively emphasized percolation criticality and cut capacity optimization, but lacks a unified framework that simultaneously accounts for spatial structure, fragmentation balance, and computational scalability. Recent reviews indicate that network type and density significantly affect dismantling algorithm performance \cite{Wandelt2018,Musciotto2022,Bellingeri2023,Sun2023}: sparse networks benefit more from balanced fragmentation strategies, whereas highly dense networks rely more on minimizing cut capacity. Against this background, our proposed dual-path budget-adaptive algorithm based on Wilson UST approximates the original graph’s connectivity structure with a spanning tree skeleton, and employs “density-informed” small-budget and “slope-prior” large-budget mechanisms to adapt to different budget scales. This framework achieves a unification of structural interpretability, balanced fragmentation, and computational efficiency while maintaining near-linear complexity, providing a new scalable approach for planar network edge removal planning and spatial robustness analysis.

\section{Methodology}
\subsection{Problem Definition}
We address the “worst-case planar graph disruption” problem, formalized as the following network dismantling model: given an undirected graph $G=(V,E)$ and an edge budget $cut$, the goal is to remove at most $cut$ edges such that the size of the largest connected component of the remaining graph is minimized:
\[
\min_{F \subseteq E,\; |F| \le cut} \; L(G \setminus F),
\] 
where $L(\cdot)$ denotes the size of the largest connected component and $F$ is the set of removed edges. Unlike the traditional minimum cut problem, which focuses on minimizing cut capacity, this problem emphasizes balanced fragmentation of the network \cite{Braunstein2016}. Since this optimization is NP-hard on general planar graphs, direct solution is infeasible. Therefore, we adopt a layered approach: first, we study a related subproblem that is more tractable and can provide quality guarantees; then, guided by the subproblem’s solution, we derive a feasible near-optimal solution for the original problem.

\subsection{Subproblem Formulation}
We introduce a subproblem as a proxy for the original problem. Given a connected graph $G=(V,E)$ and a target number of partitions $k$, consider partitioning the graph into $k$ connected components under the constraint of balanced component sizes. In other words, we aim to fragment the graph into $k$ parts such that the largest component is as small as possible, while also considering the number of cut edges. This subproblem can be approached on a spanning tree skeleton of the graph, which provides a simplified structure for designing efficient cuts. We first solve this subproblem and then use its solution as a basis to tackle the original problem.

\subsection{Subproblem Solution}

\begin{enumerate}[label=(\arabic*), leftmargin=1.5em, itemsep=1em]

\item \textbf{Uniform Spanning Tree Sampling.} We use Wilson’s algorithm to generate a uniform random spanning tree (UST) of the graph. The core of Wilson’s algorithm is performing a loop-erased random walk (LERW) from each new starting node not yet in the tree. Specifically:
\begin{enumerate}[label=(\alph*), itemsep=0pt]
    \item Pick an arbitrary root node $r$ and initialize the tree $T = \{r\}$.
    \item For any node $u$ not yet in $T$: (a) start a simple random walk from $u$ until it hits a node in $T$; (b) erase any loops in the walk path to get a simple path $P$; (c) add all edges of $P$ to $T$.
    \item Repeat until all nodes are included in $T$. The result is a uniform spanning tree of $G$ (each spanning tree of $G$ is equally likely to be produced).
\end{enumerate}
\textit{Properties and Complexity:} The expected runtime of Wilson’s algorithm on a general graph is related to the graph’s effective resistance, but for sparse or planar graphs and with practical optimizations (using adjacency lists, hit tables, and early stop), it runs in near-linear time (approximately $O(|E|)$ in our implementation). This makes it a low-cost skeleton generator suitable for multiple sampling runs.

\item \textbf{One-time Balanced Partition on the Tree.} We perform a one-time balanced bi-partition (or multi-partition) on the UST $T$: we cut $(k - 1)$ tree edges such that the resulting $k$ subtrees are as equal in size as possible. The procedure consists of four steps:

\begin{enumerate}[label=(\alph*), itemsep=0.8em]
    \item \textit{DFS Preprocessing.} Build an adjacency list for $T$ and run a depth-first search (DFS) from a root (e.g., the node with smallest ID). Record for each node $u$: its parent $parent(u)$, the edge to its parent $parent\_edge(u)$, its list of children $children(u)$, and the DFS post-order list $post$. Also compute the size of each subtree $subtree(u)$ via a post-order traversal.
    
    \item \textit{Target Share Sequence.} Let $n = |V|$. Compute $q = \lfloor n/k \rfloor$ and $r = n - qk$. Construct a target partition size list: $targets = [q+1, \dots, q+1$ ($r$ times), $q, \dots, q$ ($k-r$ times)$]$. This represents the ideal sizes of each of the $k$ components. Initialize a pointer $tPtr = 1$ for the current target index.
    
    \item \textit{Post-order Greedy Cutting.} Traverse nodes in post-order. For each node $u$, let $rem = 1 + \sum_{v \in children(u)} subtree(v)$. If $u$ is not the root, fewer than $(k - 1)$ cuts have been made so far, and $rem \ge targets[tPtr]$, then cut the edge to $parent(u)$, record the component size, and increment $tPtr$. Set $rem = 0$ for that subtree. Continue until all nodes are processed.
    
    \item \textit{Post-cut Adjustment.} If fewer than $(k - 1)$ cuts have been made (which can happen in special cases like a chain-like tree), perform additional cuts greedily based on which remaining subtree sizes are closest to the remaining target sizes. This ensures the total of $(k - 1)$ cuts is reached. (This step has cost $O((k - 1) \log n)$.)
\end{enumerate}

After these steps, we obtain the final set of cut edges $F$ and the partition set $P$. The complexity of this balanced partition procedure is: DFS and subtree size computation: $O(n)$; post-order greedy partitioning: $O(n)$; adjustment phase: $O(n)$ in worst case; overall approximately linear in $|V|$. Pseudocode is given in Algorithm~\ref{alg:balanced}.

\begin{algorithm}[ht]
\caption{One-time Balanced Partition (post-order greedy)}\label{alg:balanced}
\KwIn{Spanning tree $T$, target number of partitions $k$.}
\KwOut{Cut edge set $F$, partition set $P$.}
Perform a single-stack DFS to compute $parent(\cdot)$, $parent\_edge(\cdot)$, $children(\cdot)$, post-order list $post$, and subtree sizes $subtree(\cdot)$\;
Compute $n = |V|$, $q = \lfloor n/k \rfloor$, $r = n - qk$, and $targets = [\underbrace{q+1,\dots,q+1}_{r \text{ times}}, \underbrace{q,\dots,q}_{k-r \text{ times}}]$; set $tPtr \leftarrow 1$\;
\For{$u \in post$}{
    $rem \leftarrow 1 + \sum_{v \in children(u)} subtree(v)$\;
    \If{$u$ is not root \textbf{and} cuts made $< k-1$ \textbf{and} $rem \ge targets[tPtr]$}{
        cut $parent\_edge(u)$; record component size; $tPtr \leftarrow tPtr + 1$\;
        $rem \leftarrow 0$\;
    }
}
\If{\#cuts $< k-1$}{
    (Post-cut phase) Greedily cut remaining edges closest to targets\;
}
\Return $F$, $P$\;
\end{algorithm}

\item \textbf{Result Evaluation after Multiple Trials.} We repeat the “UST sampling + tree partition” experiment $m$ times with different random seeds (using a seed list $\{s_1, s_2, \dots, s_m\}$). After each trial, we record the resulting partition and its LCC. We then select the best result under a specified rule. We consider two criteria:  
\textbf{Criterion A:} the smallest LCC size achieved;  
\textbf{Criterion B:} the fewest cut edges $|F|$ among solutions with the same LCC (tie-breaker).  
Depending on the problem requirements, we prioritize Criterion A and then Criterion B to choose the optimal solution. For example, one decision rule could be: (1) primarily select the solution with minimum LCC; (2) if multiple solutions have the same LCC, choose the one with fewer cut edges. Following this rule (corresponding to a parameter $\lambda$ in the problem model controlling priority), we pick the best solution $(F^*, P^*)$ from all trials.

\subsubsection*{Subproblem Complexity and Parallelism}

The time complexity for a single “sampling + partition” trial is approximately $O(|E|)$ (for UST generation) + $O(|V|)$ (for tree partition), which is about $O(|E|)$ since the graphs are sparse. For $m$ trials, the total cost is $O(m\,|E|)$. The final evaluation stage (comparing $m$ results) is $O(m)$. Overall, the subproblem solver runs in $O(m\,|E|)$ time. Moreover, since trials with different random seeds can be executed independently, this approach is easily parallelizable across multiple threads or GPUs.

\end{enumerate}

\subsection{Integrated Framework for the Original Problem}
Let the density feature $t = \log|E|/\log|V|$. When $k = 2$, the optimal cut set (and its cost) for the subproblem is denoted $F^{(2)} = F^*(k{=}2)$, with $|F^{(2)}|$ edges. Using the $k=2$ subproblem solution as a baseline, and given the graph’s density $t$ and the budget $cut$, we construct two paths to solve the original problem:
\begin{enumerate}\itemsep0pt
    \item If $cut < |F^{(2)}|$: Estimate the disruptable subgraph size fraction $\alpha$, extract a subgraph $G_s$ with $\alpha n$ nodes from $G$, and solve a $k=2$ partition on $G_s$; finally map the result back to $G$ to form the overall cut solution.
    \item If $cut \ge |F^{(2)}|$: Estimate the optimal number of partitions $\hat{k}$ and perform a search over partitions around $\hat{k}$ on the full graph; output the solution that minimizes LCC within the budget.
\end{enumerate}

\textbf{Small-budget path ($cut < |F^{(2)}|$): Estimating $\alpha$ and subgraph partition.} When the edge budget is insufficient to achieve the baseline bisection (i.e., cannot remove $|F^{(2)}|$ edges), we first estimate $\alpha \in (0,1)$, the fraction of nodes to dismantle. We use an empirical regression of the form $\hat{\alpha} = g_{\theta}(cut, t, |F^{(2)}|)$ (for example, $\hat{\alpha} = B_0\, t^{-\gamma} \cdot (\textit{cut})^{\beta_1} \cdot |F^{(2)}|^{-\beta_2}$) fitted from preliminary experiments. Then we construct a subgraph $G_s$ by selecting $n_s = \lfloor \hat{\alpha}\, n \rfloor$ nodes. The selection strategy starts from an initial corner of the network and iteratively adds the frontier node with the smallest degree, ensuring we obtain a relatively sparse, localized subgraph of size $n_s$. We then solve the $k=2$ subproblem on $G_s$, and finally map the resulting cut edges back to the original graph to form the overall solution.

\textbf{Large-budget path ($cut \ge |F^{(2)}|$): Estimating $\hat{k}$ and searching $k$-partitions.} When the budget is sufficient to support a balanced multi-way partition of the graph, empirical observation suggests an approximately linear relationship between optimal cut size and number of partitions: $|F^{(k)}| \approx |F^{(2)}| + s(t)\,(k-2)$, where $s(t)$ indicates the average additional cut cost per extra partition. Based on this, we estimate an initial $\hat{k}_0 = \max\{2, \text{round}(2 + (cut - |F^{(2)}|)/s(t))\}$. Allowing for model noise, we then search in a small window around $\hat{k}_0$ (e.g., $\hat{k}_0 \pm \Delta$) and for each $k$ in this set, run the subproblem solver (with multiple spanning tree samples). Among all candidate solutions that do not exceed the budget, we select the one with the smallest LCC as the output. Empirically, $s(t)$ grows non-linearly with $t$; in our implementation we approximate it by a model $s(t) = c_0 \ln t + c_1$.

Algorithm~\ref{alg:integrated} outlines the integration of these steps into a unified solver for the original problem.

\begin{algorithm}[ht]
\caption{Integrated Solver for Planar Graph Disruption (given budget $cut$)}\label{alg:integrated}
\KwIn{$G=(V,E)$, budget $cut$, seed list, slope model $s(t)$, fit function $g_{\theta}$}
\KwOut{Edge removal set $F^*$, partition scheme $P^*$, metrics}
$t \leftarrow \log |E| / \log |V|$\; 
\textit{// Baseline: solve subproblem for $k=2$}\;
Compute optimal $|F^{(2)}|$ via “sample-cut-evaluate” on $G$ with $k=2$\;
\eIf{$cut < |F^{(2)}|$}{
    \textit{// Small-budget path: estimate $\alpha$ and partition subgraph}\;
    $\hat{\alpha} \leftarrow g_{\theta}(cut, t, |F^{(2)}|)$; $n_s \leftarrow \lfloor \hat{\alpha}\, |V| \rfloor$\;
    construct subgraph $G_s$ with $n_s$ nodes\;
    solve subproblem on $G_s$ with $k=2$, map result back to $G$\;
    $(F^*, P^*) \leftarrow$ solution minimizing $L(G \setminus F)$\;
}{
    \textit{// Large-budget path: estimate $\hat{k}$ and search partitions}\;
    $\hat{k}_0 \leftarrow \max\{2,\; \text{round}(2 + (cut - |F^{(2)}|)/s(t))\}$\;
    \For{$k \in \{\hat{k}_0 - \Delta, \dots, \hat{k}_0 + \Delta\}$}{
        run “sample-cut-evaluate” on $G$ for $k$ (multi-seed)\;
        record best candidate $(F_k, P_k)$ within budget\;
    }
    $(F^*, P^*) \leftarrow$ best candidate (lowest LCC)\;
}
\Return $(F^*, P^*)$ (and runtime statistics)\;
\end{algorithm}

\textit{Complexity and Parallelism.} The total computational cost of the integrated solver has three components: baseline stage, path execution, and final selection. (1) Baseline: one full “sample-cut-evaluate” run at $k=2$, complexity $\sim O(m\,|E|)$ (with $m$ samples). (2) Small-budget path (if applicable): runs on a subgraph of size $n_s = \hat{\alpha} n$, with $|E_s| \approx \hat{\alpha}|E|$, so complexity $O(m\,|E_s|) \approx O(m\, \hat{\alpha}\, |E|)$, which scales linearly with the estimated fragment size $\hat{\alpha}$. (3) Large-budget path (if applicable): after estimating $\hat{k}_0$, search over a small window $K = \{\hat{k}_0 - \Delta, \dots, \hat{k}_0 + \Delta\}$ (window size $|K| = 2\Delta+1$, usually constant), yielding complexity $O(m\, |E|\, |K|)$. Thus, the overall complexity can be expressed as $T_{\text{total}} = O(m\,|E|) + O(m\, \hat{\alpha}\, |E|) + O(m\, |E|\, |K|)$, dominated by $O(m\, |E|\, |K|)$.

Because different random seed experiments and different $k$ values are independent, the integrated algorithm is naturally parallelizable. In a multi-core CPU or GPU environment, we can parallelize across the “seed dimension” and the “partition dimension” to achieve near-linear speedup. The parallel execution time is approximately $T_{\text{parallel}} \approx O(m\,|E|\,|K|)/p$, where $p$ is the number of parallel threads/devices. Therefore, the integrated framework maintains near-linear complexity while efficiently scaling to worst-case disruption analysis on large planar graphs.

\section{Experiments and Results}
\subsection{Experimental Setup}
To evaluate the effectiveness and scalability of the proposed spanning-tree-based dual-path dismantling algorithm, we conduct experiments on two sets of random planar network datasets of different scales:
\begin{itemize}\itemsep0pt
    \item \textbf{Small-scale dataset (RandomPlanarNetworks\_200):} Randomly generated connected planar graphs with $|V|=200$ nodes and edge counts $|E|$ ranging from 250 to 550.
    \item \textbf{Large-scale dataset (PlanarNetwork\_N10000\_E11000-29000):} Randomly generated connected planar graphs with $|V| = 10^4$ nodes and $|E|$ ranging from $1.1\times 10^4$ to $2.9\times 10^4$.
\end{itemize}
All experiments were conducted on a workstation with an NVIDIA RTX 5000 Ada (32GB) GPU and an Intel i9-13950HX CPU (128GB RAM), using MATLAB R2024b. Key algorithm parameters were set as: number of spanning tree samples $m = 50$; partition search window $\Delta k = 1$; density feature $t = \log|E|/\log|V|$; slope model $s(t) = c_0 \ln t + c_1$. We evaluate budget fractions $cut/|E|$ from 5\% to 20\%.

\subsection{Evaluation Metrics}
We use three evaluation metrics:
\begin{enumerate}\itemsep0pt
    \item \textbf{LCC size ratio $L_{\max}(G \setminus F)/|V|$:} the proportion of nodes in the largest connected component after edge removals, measuring residual connectivity.
    \item \textbf{Per-edge efficiency $\eta = \frac{L_{\max}(G) - L_{\max}(G \setminus F)}{|F|}$:} the reduction in LCC per removed edge (higher is better).
    \item \textbf{Average runtime $T$:} the mean running time of the algorithm.
\end{enumerate}

\subsection{Results and Analysis}
Overall, the framework operates in three steps: (1) perform a planar graph bisection based on the spanning tree skeleton to obtain the cost baseline $|F^{(2)}|$; (2) compare the given budget $cut$ with $|F^{(2)}|$ to determine which algorithm path to follow; (3) execute the subsequent cuts and fragmentation on the chosen path. We present results illustrating each step and the performance of the two paths.

\subsubsection*{Basic Planar Graph Bisection}
\setlength{\belowcaptionskip}{-8pt}  

\begin{figure}[ht]
  \centering
  \begin{subfigure}[t]{0.32\linewidth}
    \centering
    \includegraphics[width=\linewidth]{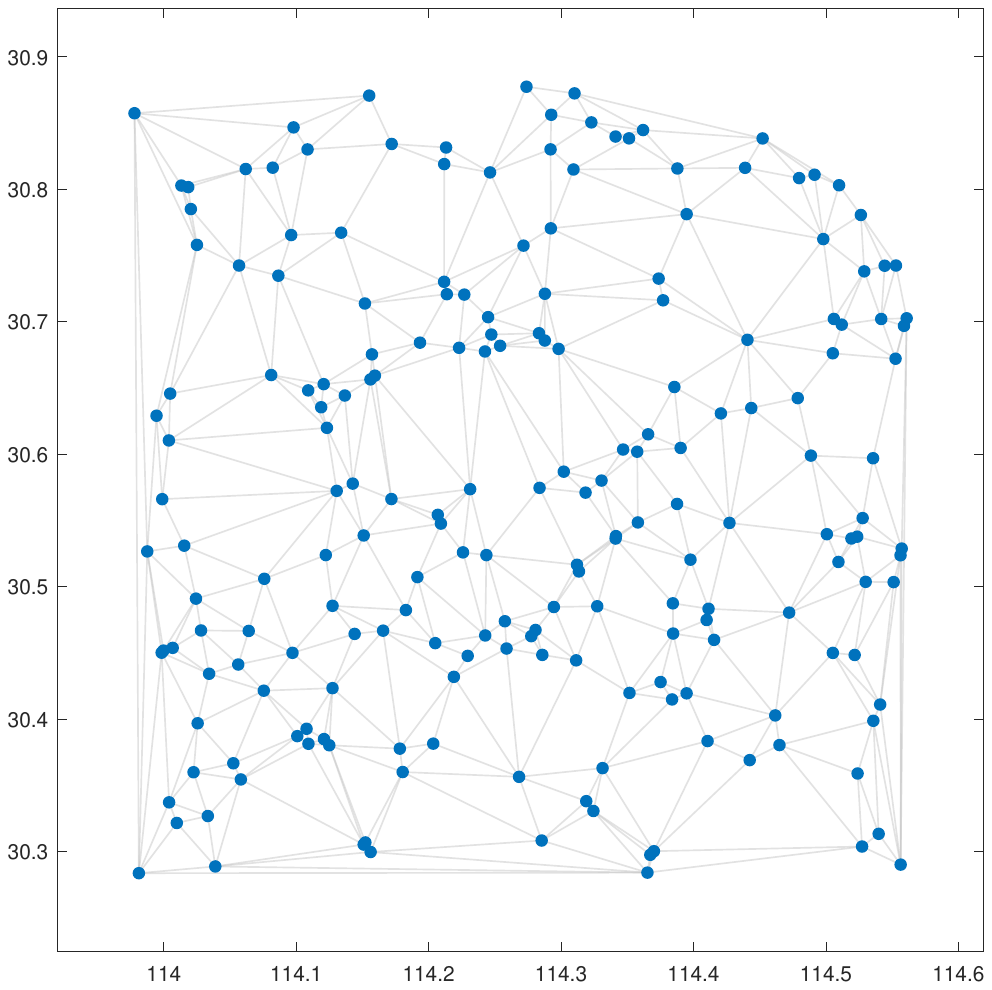}
    \caption{Original graph $G(V,E)$}
  \end{subfigure}
  \hfill
  \begin{subfigure}[t]{0.32\linewidth}
    \centering
    \includegraphics[width=\linewidth]{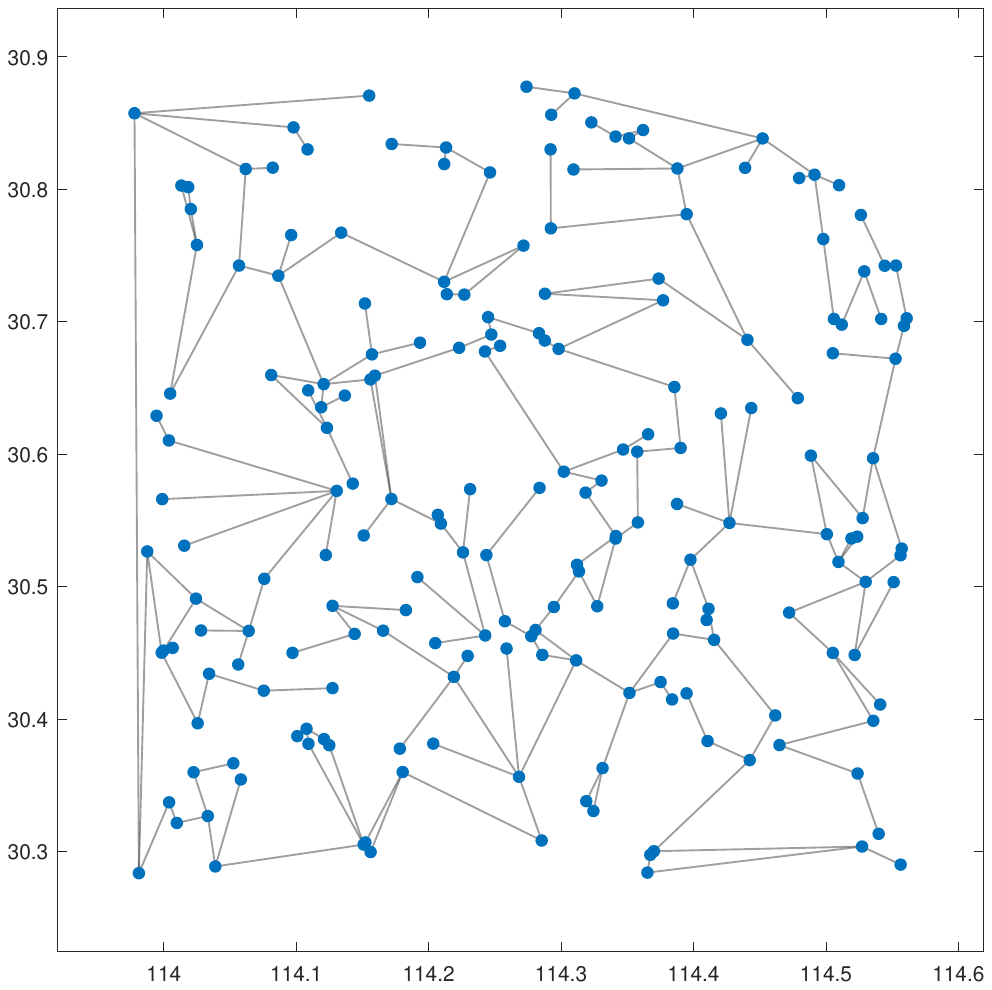}
    \caption{Spanning tree sampling and skeleton extraction}
  \end{subfigure}
  \hfill
  \begin{subfigure}[t]{0.32\linewidth}
    \centering
    \includegraphics[width=\linewidth]{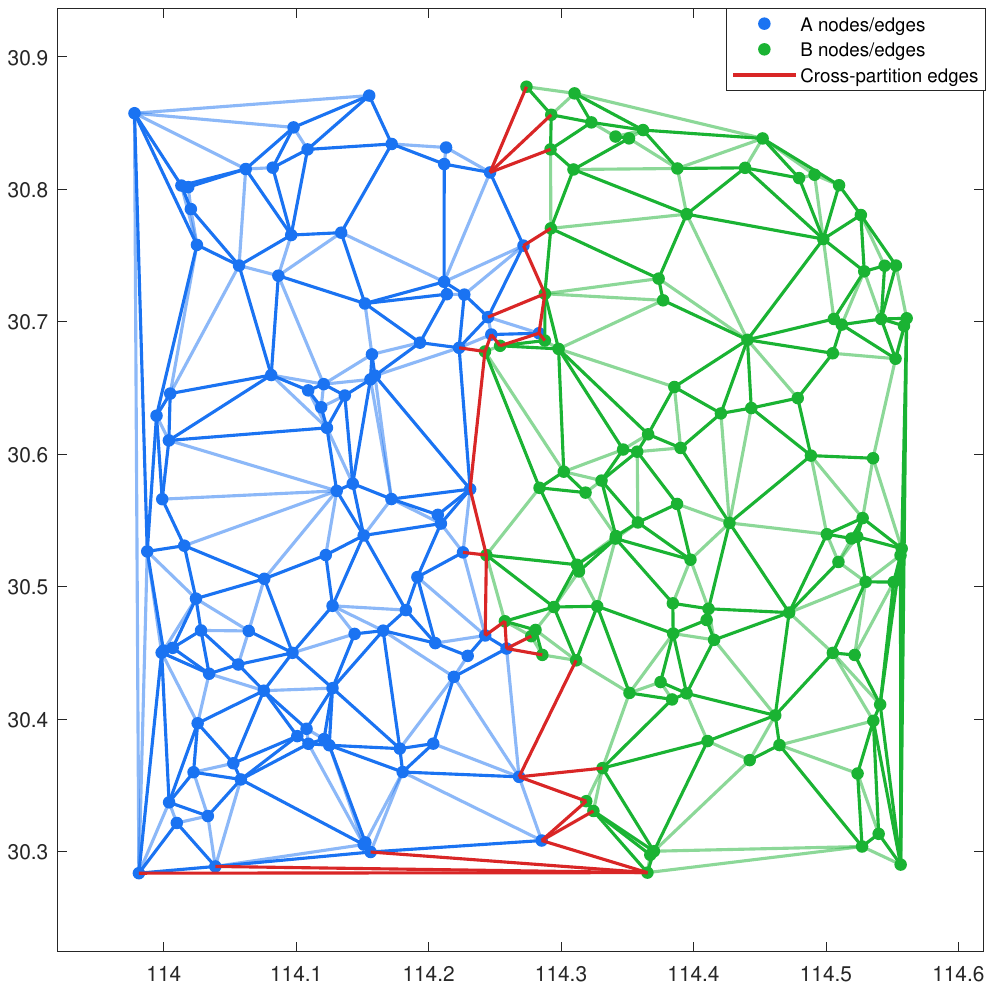}
    \caption{Balanced bisection result}
  \end{subfigure}

  \vspace{-1.0em}  
  \caption{Illustration of planar graph bisection based on a spanning tree skeleton. (a) The original graph $G(V,E)$ with 200 nodes and 550 edges. (b) A spanning tree skeleton sampled from the original graph. (c) The balanced bisection result. The spanning-tree-based planar bisection yields a baseline cut cost $|F^{(2)}|$, which, together with the given budget $cut$, determines the subsequent algorithm path.}
  \label{fig:dad_process_basic}
\end{figure}

\subsubsection*{Small-Budget Path (Estimation and Subgraph Partition)}
When the budget ratio satisfies $cut < |F^{(2)}|$, the algorithm follows the small-budget path: it estimates the disruptable subgraph size and performs a bipartition on the spanning-tree-induced subgraph. The key idea is to use the density feature $t = \log|E|/\log|V|$ and the empirical function $g_{\theta}$ to predict the fraction of the network $\alpha$ that can be dismantled, thereby determining the size of the subgraph to weaken and the target number of edges to cut.

\begin{figure}[H]
  \centering
  \begin{subfigure}[t]{0.45\linewidth}
    \centering
    \includegraphics[width=\linewidth]{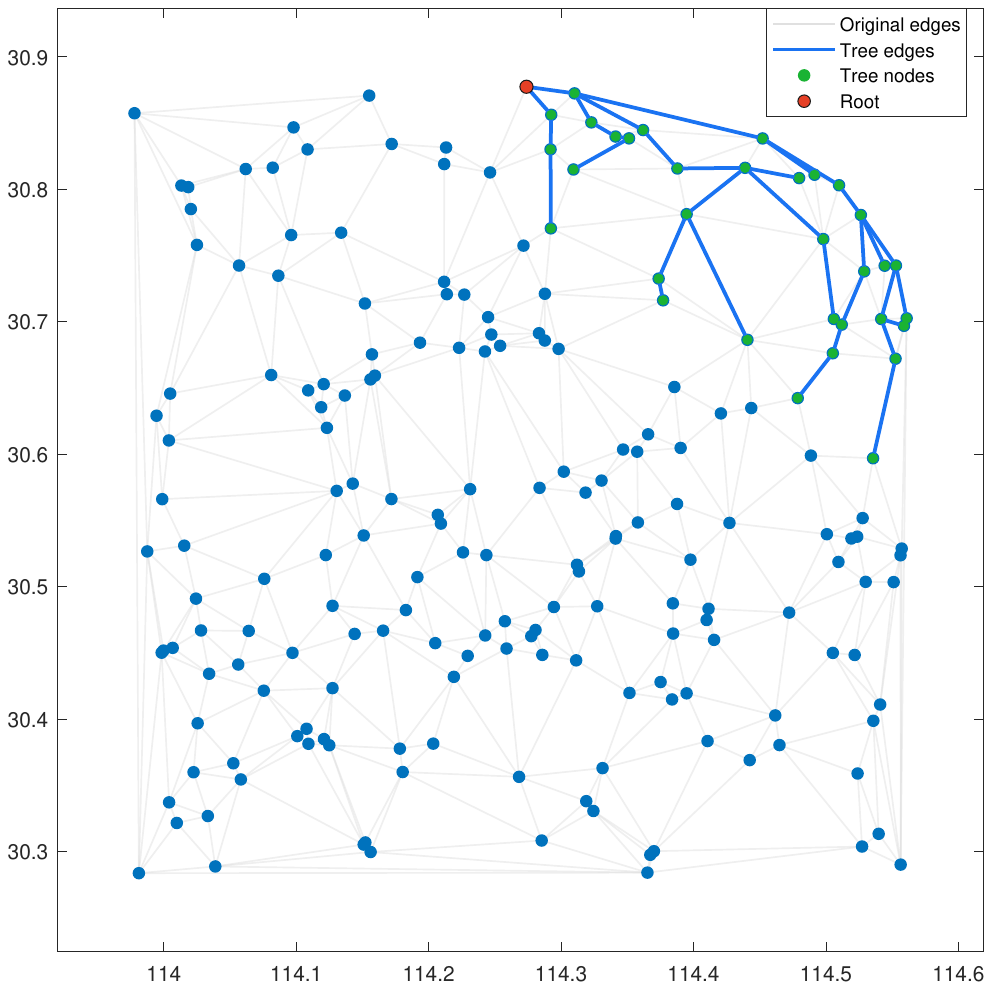}
    \caption{Spanning tree under budget-induced constraint}
  \end{subfigure}
  \hfill
  \begin{subfigure}[t]{0.45\linewidth}
    \centering
    \includegraphics[width=\linewidth]{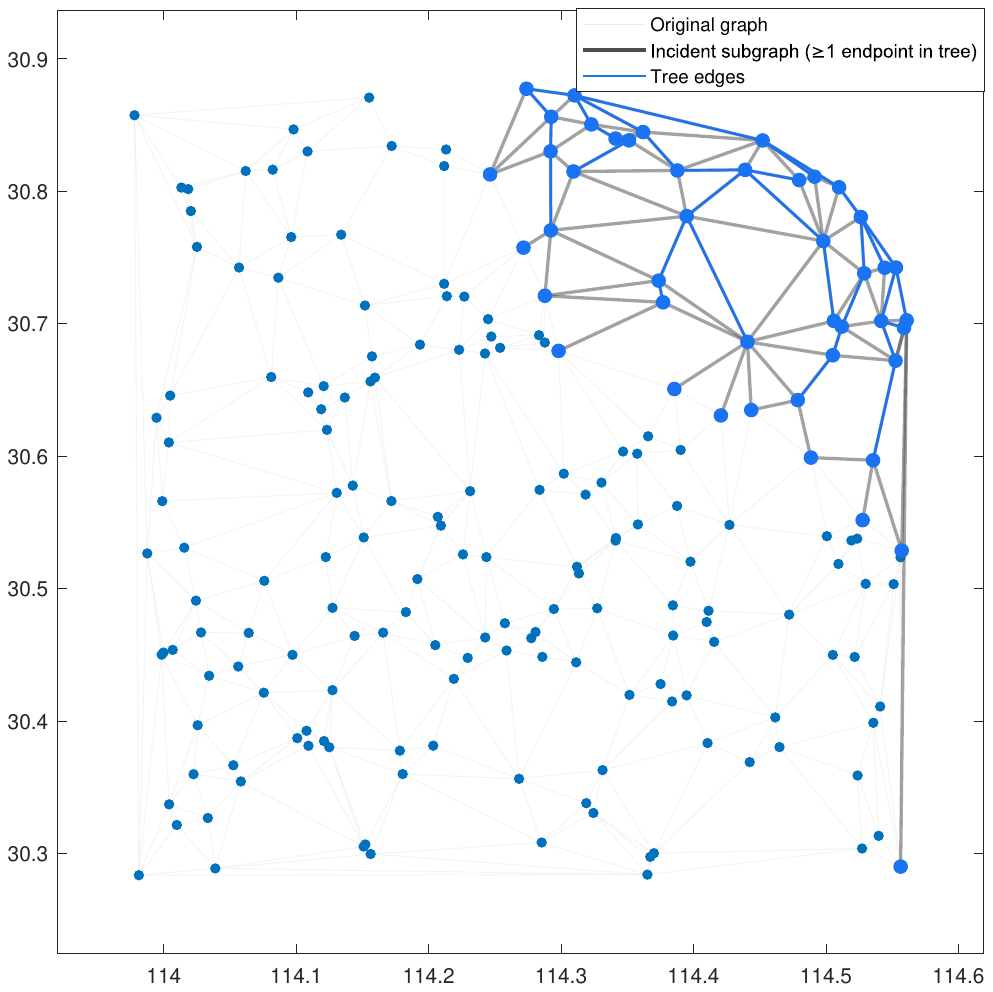}
    \caption{Skeleton-induced subgraph}
  \end{subfigure}\\[-1.2ex]  

  \begin{subfigure}[t]{0.45\linewidth}
    \centering
    \includegraphics[width=\linewidth]{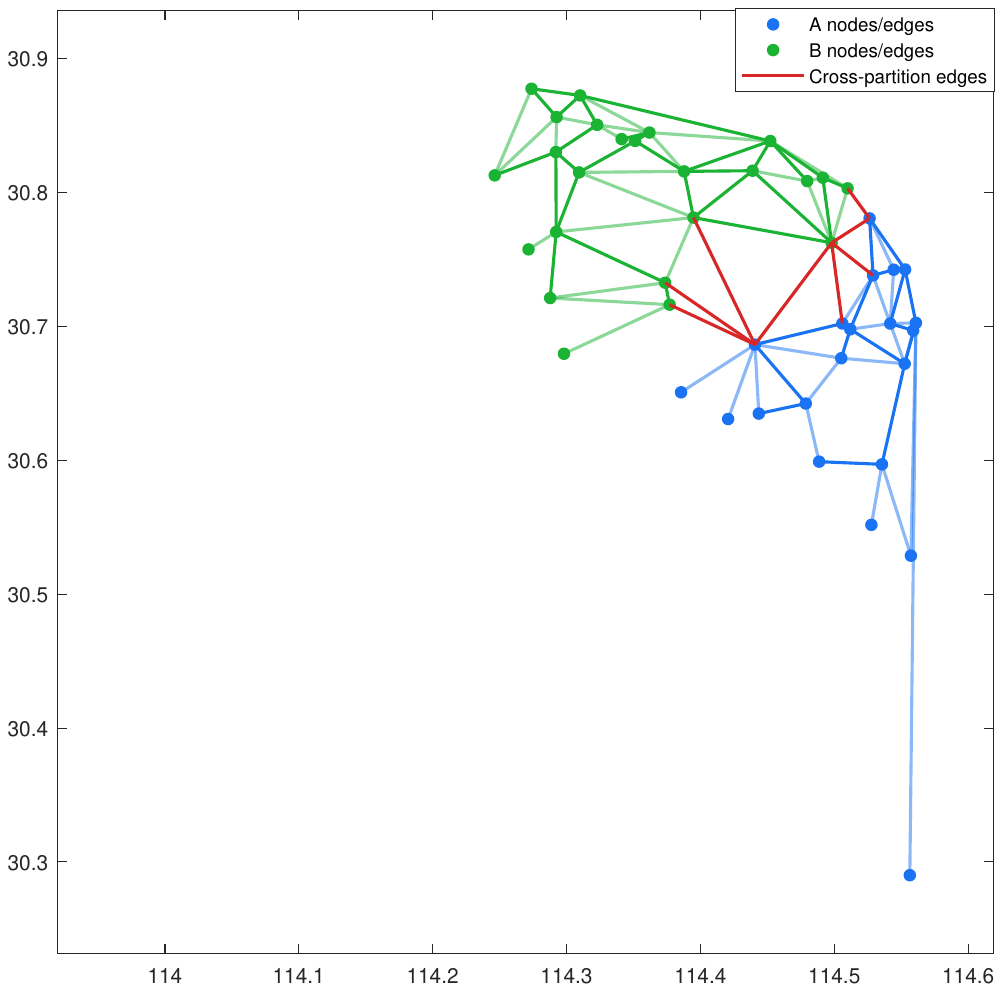}
    \caption{Subgraph bipartition result}
  \end{subfigure}
  \hfill
  \begin{subfigure}[t]{0.45\linewidth}
    \centering
    \includegraphics[width=\linewidth]{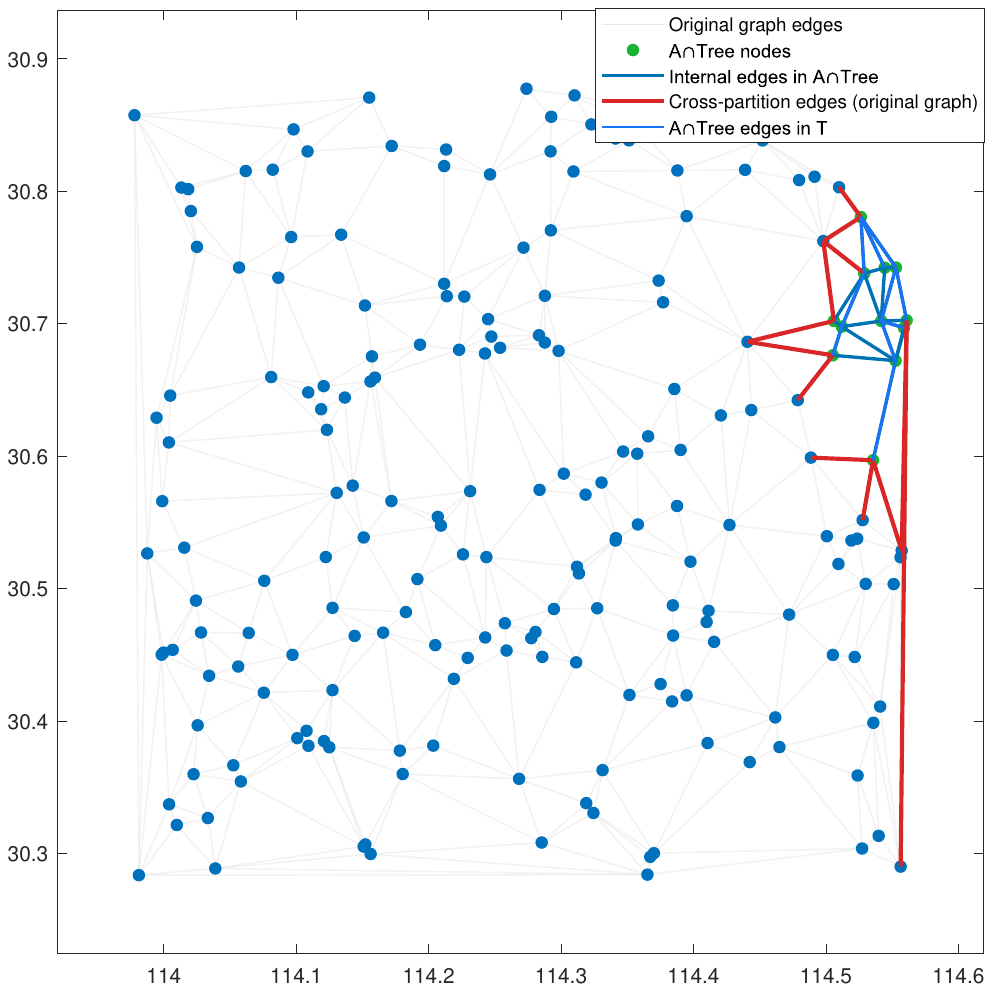}
    \caption{Final cut result on original graph}
  \end{subfigure}

  \vspace{-1.0em}
  \caption{Example of the small-budget path process. (a) A spanning tree sampled under the estimated fraction $\hat{\alpha}$. (b) The subgraph induced by the spanning tree skeleton. (c) The bipartition result on the subgraph. (d) The corresponding partitioning result mapped onto the original graph. This process involves only a single spanning tree and local subgraph operations, yielding a fragmentation outcome quickly with complexity on the order of $O(|E|)$.}
  \label{fig:smallbudget_demo}
\end{figure}

\subsubsection*{Large-Budget Path (Global $k$-Partitioning)}
When $cut \ge |F^{(2)}|$, the algorithm enters the large-budget path: it adaptively estimates $\hat{k}$ (the number of partitions) and performs a $k$-partition on the full graph with cross-partition edge optimization, to achieve more extensive and balanced fragmentation.

\begin{figure}[H]
  \centering
  \begin{subfigure}[t]{0.30\linewidth}
    \centering
    \includegraphics[width=\linewidth]{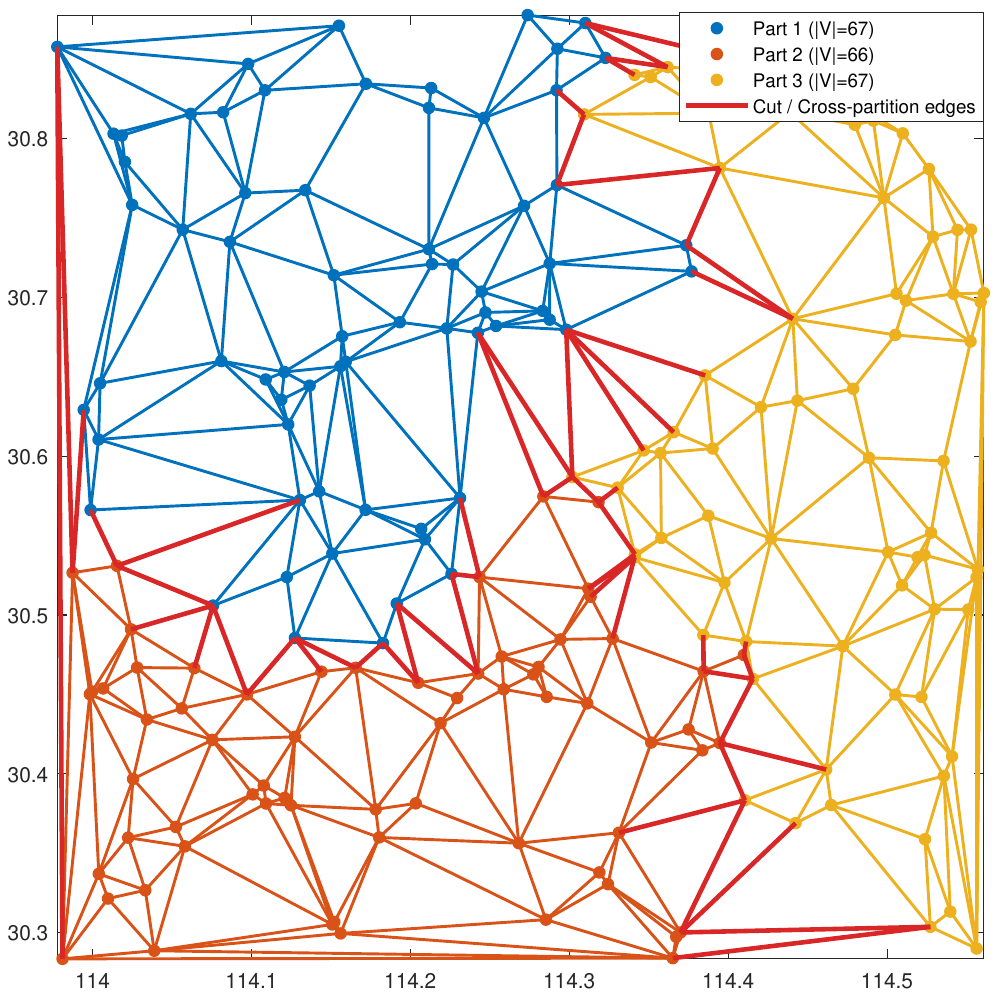}
    \caption{$k=3$ partition result}
  \end{subfigure}
  \hfill
  \begin{subfigure}[t]{0.30\linewidth}
    \centering
    \includegraphics[width=\linewidth]{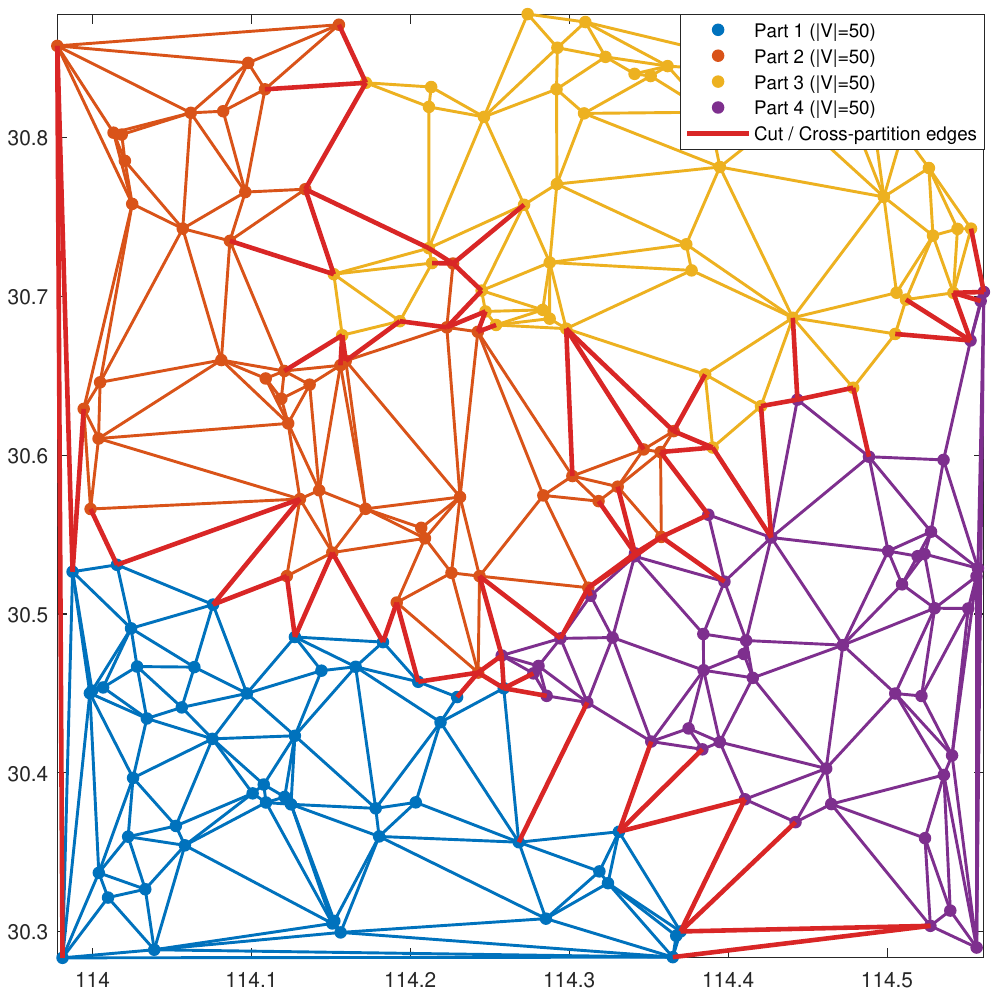}
    \caption{$k=4$ partition result}
  \end{subfigure}
  \hfill
  \begin{subfigure}[t]{0.30\linewidth}
    \centering
    \includegraphics[width=\linewidth]{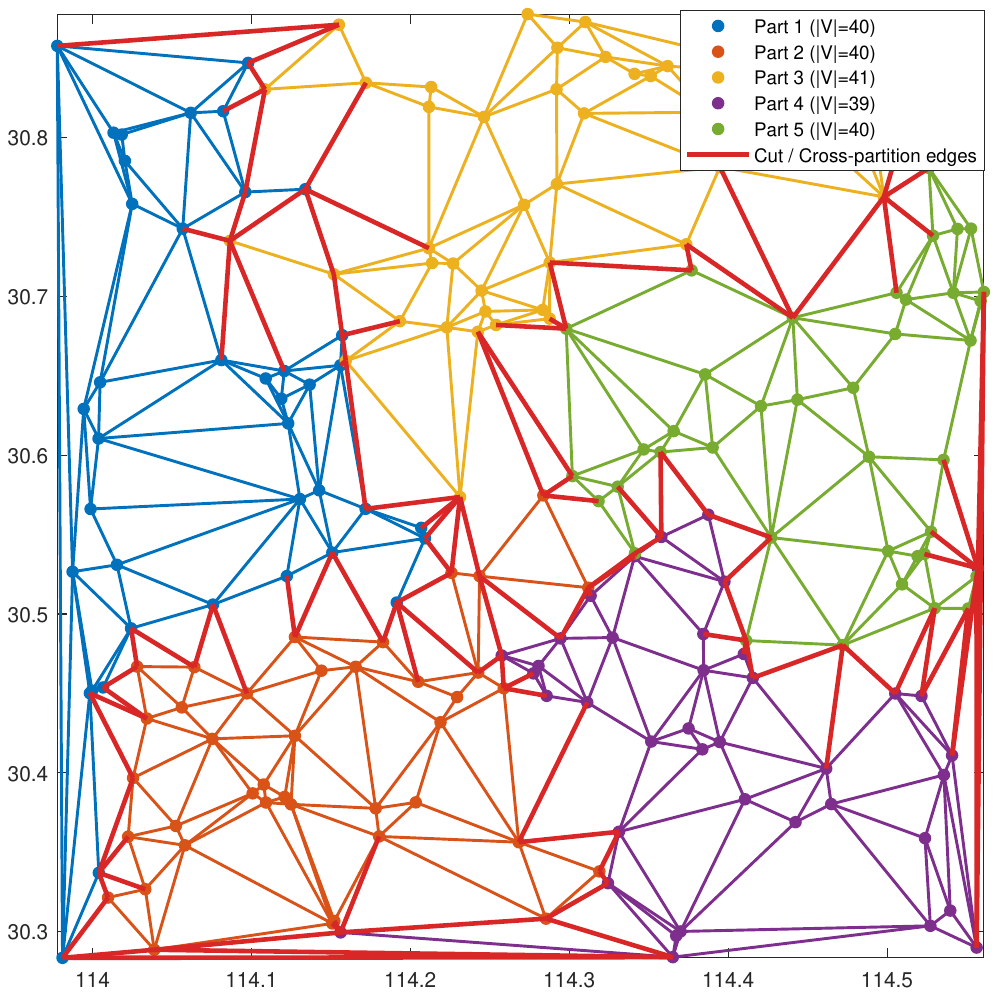}
    \caption{$k=5$ partition result}
  \end{subfigure}
  \caption{Partition examples on a small-scale planar graph under different target partition counts. As $k$ increases, the graph is split into more balanced, smaller components, and the LCC decreases.}
  \label{fig:planar_partition_examples}
\end{figure}

\textbf{Small-scale dataset results.} To illustrate the spatial fragmentation patterns for different values of $k$ in a small-scale network, Figure~\ref{fig:planar_partition_examples} presents examples of partitions on the same planar graph for $k = 3, 4, 5$. We observe that as $k$ increases, the number of cut edges rises and the original graph is broken into finer subcomponents; meanwhile, the largest connected component $L_{\max}$ gradually decreases and fragmentation becomes more balanced across regions.

\begin{figure}[H]
  \centering
  \includegraphics[width=0.78\linewidth]{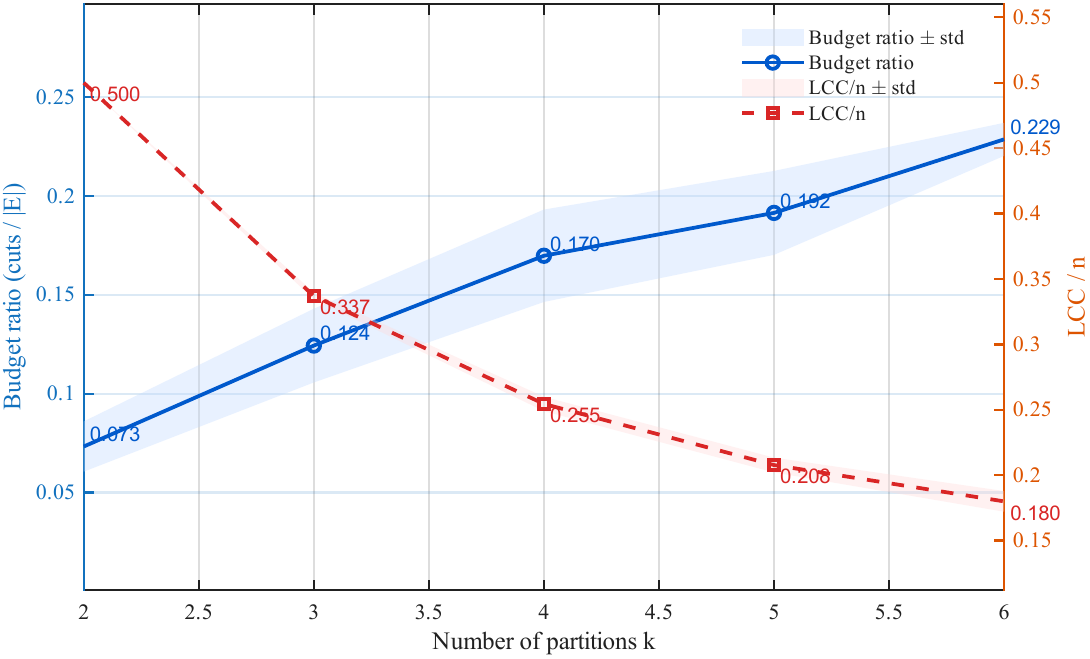}
  \caption{Small-scale dataset results – joint trends of budget ratio ($cut/|E|$, left axis) and LCC size ratio ($L_{\max}/|V|$, right axis) as $k$ increases. Curves represent means across random seeds; shaded areas show standard deviation.}
  \label{fig:k_budget_lcc_small}
\end{figure}

Figure~\ref{fig:k_budget_lcc_small} shows the combined trend of budget ratio vs. LCC ratio as $k$ varies in the small-scale dataset. The horizontal axis is $k$, the left vertical axis is $cut/|E|$, and the right vertical axis is $L_{\max}/|V|$. We see that as $k$ increases, the required budget $cut/|E|$ monotonically rises, and the LCC ratio $L_{\max}/|V|$ monotonically declines. In this small network, each $k$-partitioning has an average runtime of about 5 seconds. The curves in Figure~\ref{fig:k_budget_lcc_small} are averaged over random seeds, and the shaded bands indicate the standard deviation range, showing that the results are consistent across different runs.

\begin{figure}[H]
  \centering
  \begin{subfigure}[t]{0.45\linewidth}
    \centering
    \includegraphics[width=\linewidth]{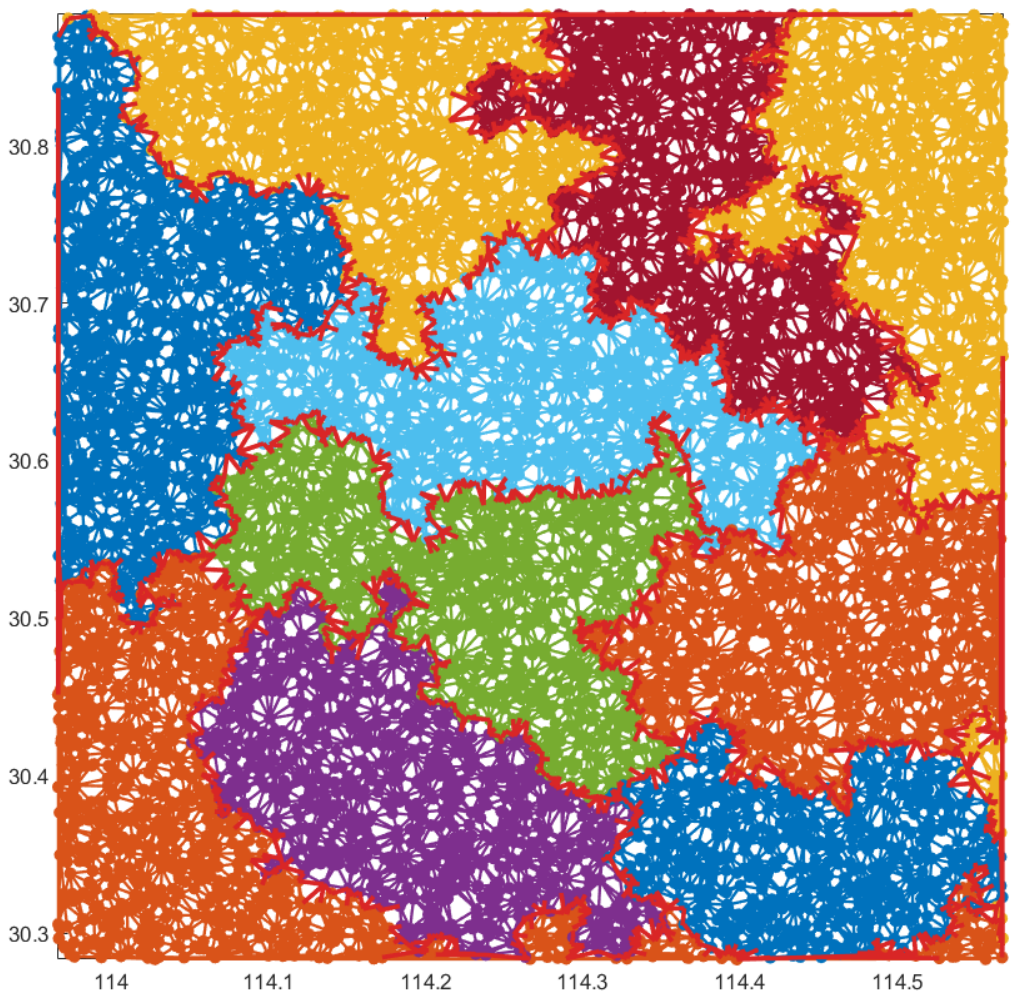}
    \caption{$k=10$ partition result}
  \end{subfigure}
  \hfill
  \begin{subfigure}[t]{0.45\linewidth}
    \centering
    \includegraphics[width=\linewidth]{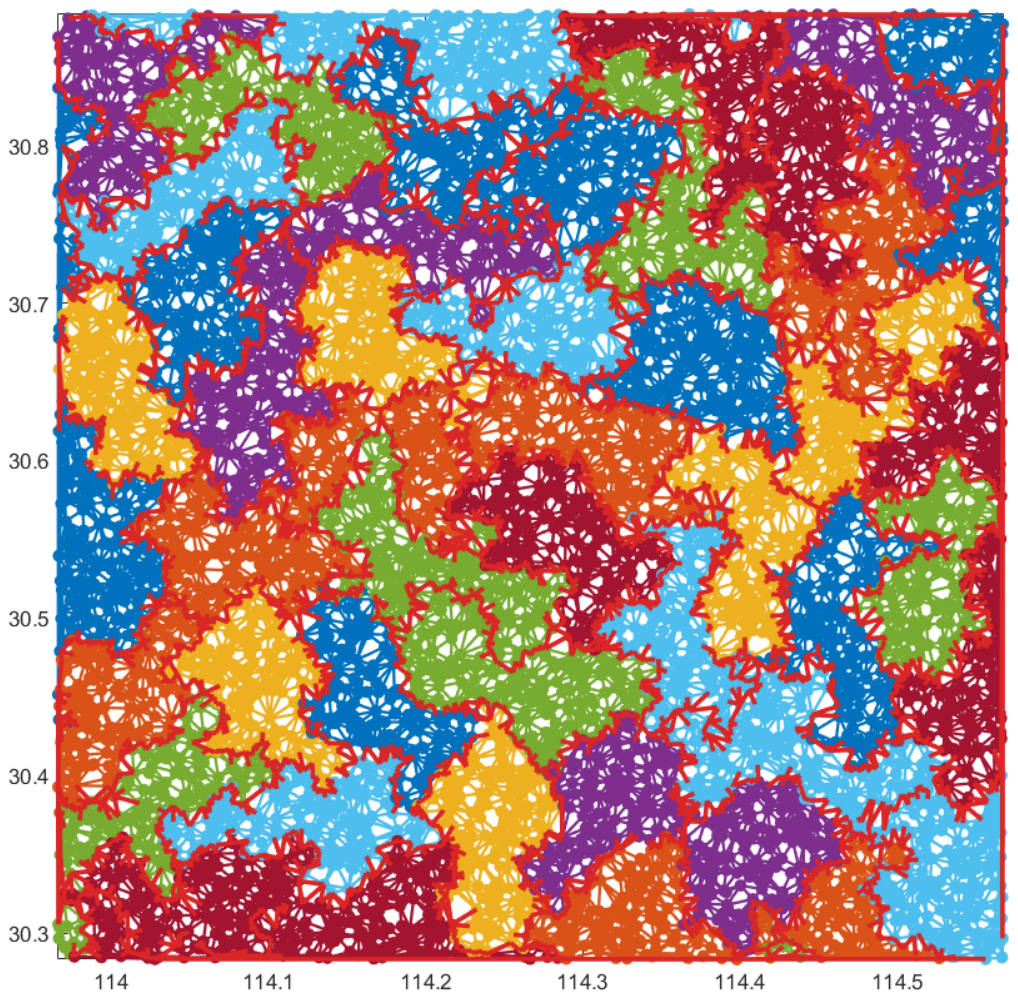}
    \caption{$k=50$ partition result}
  \end{subfigure}\\[1ex]
  \begin{subfigure}[t]{0.45\linewidth}
    \centering
    \includegraphics[width=\linewidth]{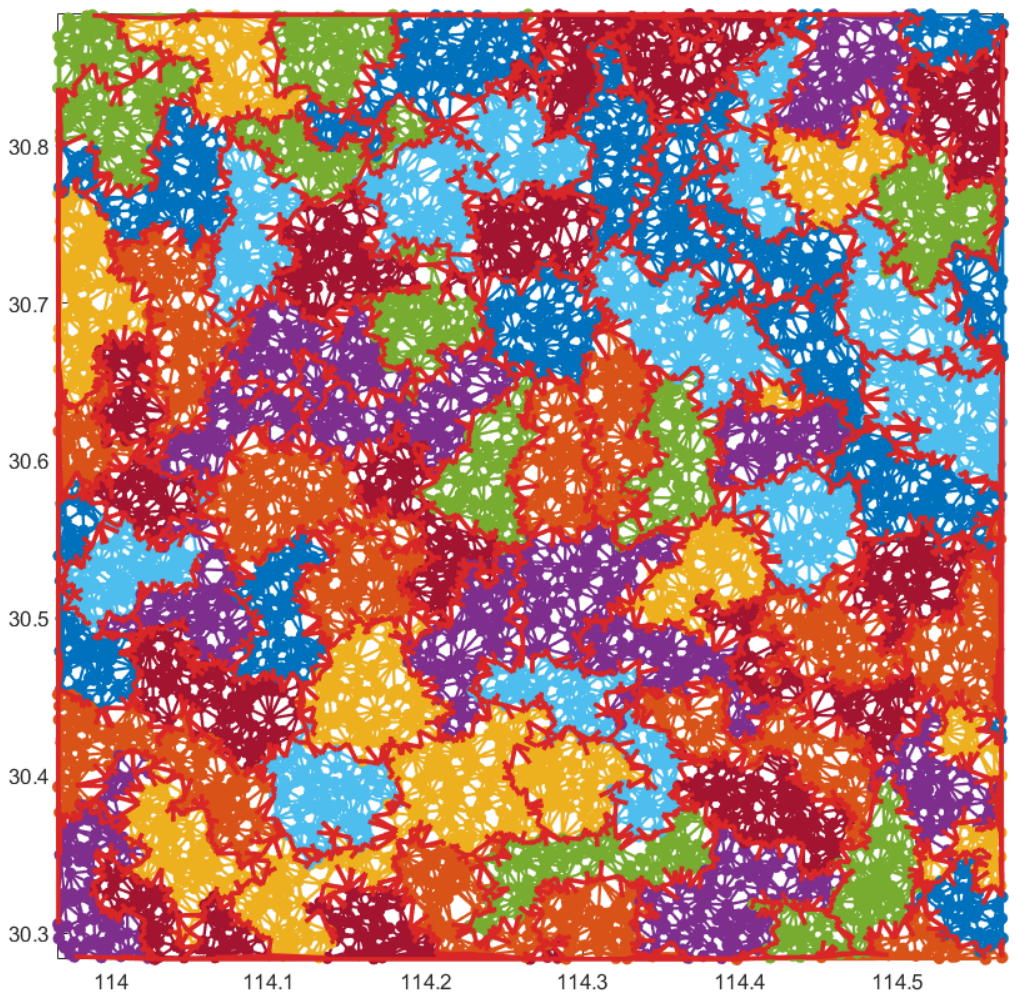}
    \caption{$k=90$ partition result}
  \end{subfigure}
  \hfill
  \begin{subfigure}[t]{0.45\linewidth}
    \centering
    \includegraphics[width=\linewidth]{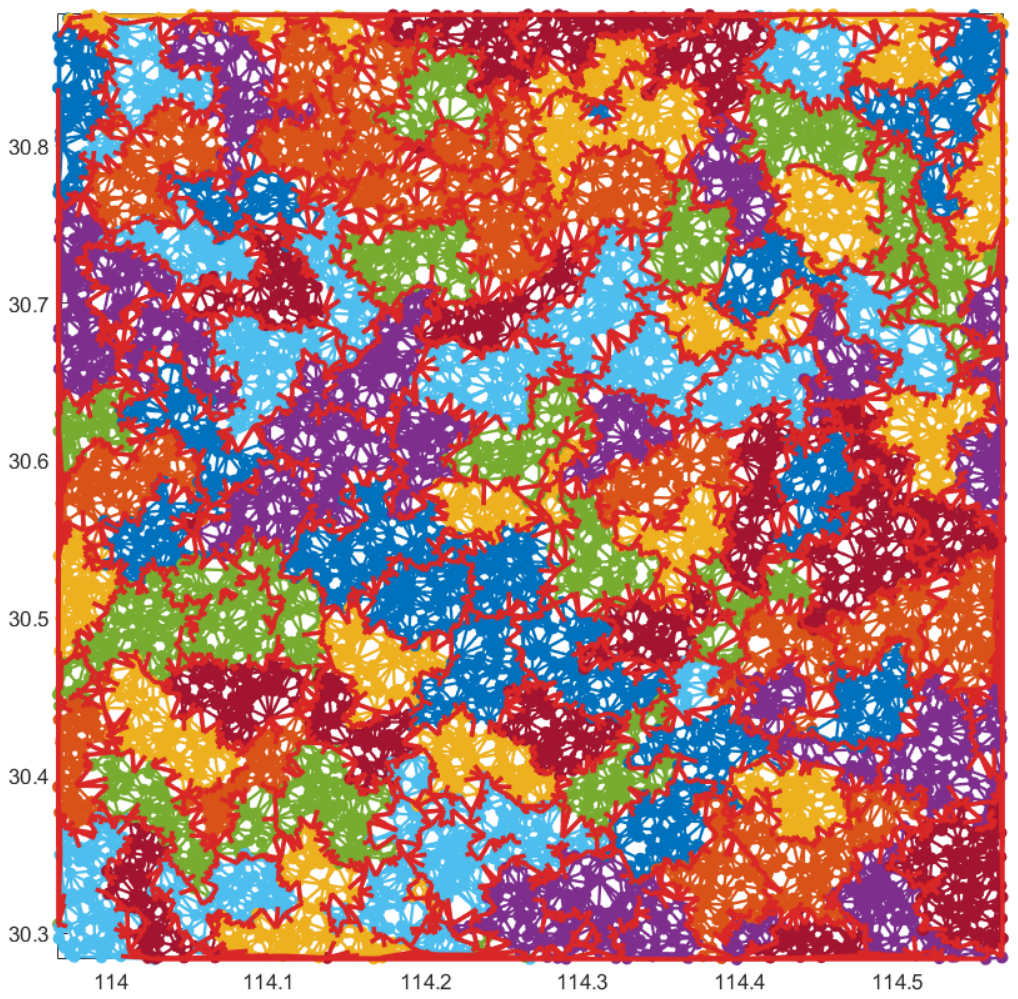}
    \caption{$k=130$ partition result}
  \end{subfigure}
  \caption{Partition examples on a large-scale planar graph for different $k$. As $k$ increases, the number of partitions grows and LCC shrinks, with balanced fragmentation across components.}
  \label{fig:big_planar_partition_examples}
\end{figure}

\textbf{Large-scale dataset results.} Similarly, to illustrate fragmentation patterns in a large network, Figure~\ref{fig:big_planar_partition_examples} shows partition examples on a large planar graph for $k = 10, 50, 90, 130$. As in the small network, increasing $k$ yields more cut edges and finer partitions, with the largest component $L_{\max}$ decreasing and fragmentation remaining balanced across components.

\begin{figure}[H]
  \centering
  \includegraphics[width=0.78\linewidth]{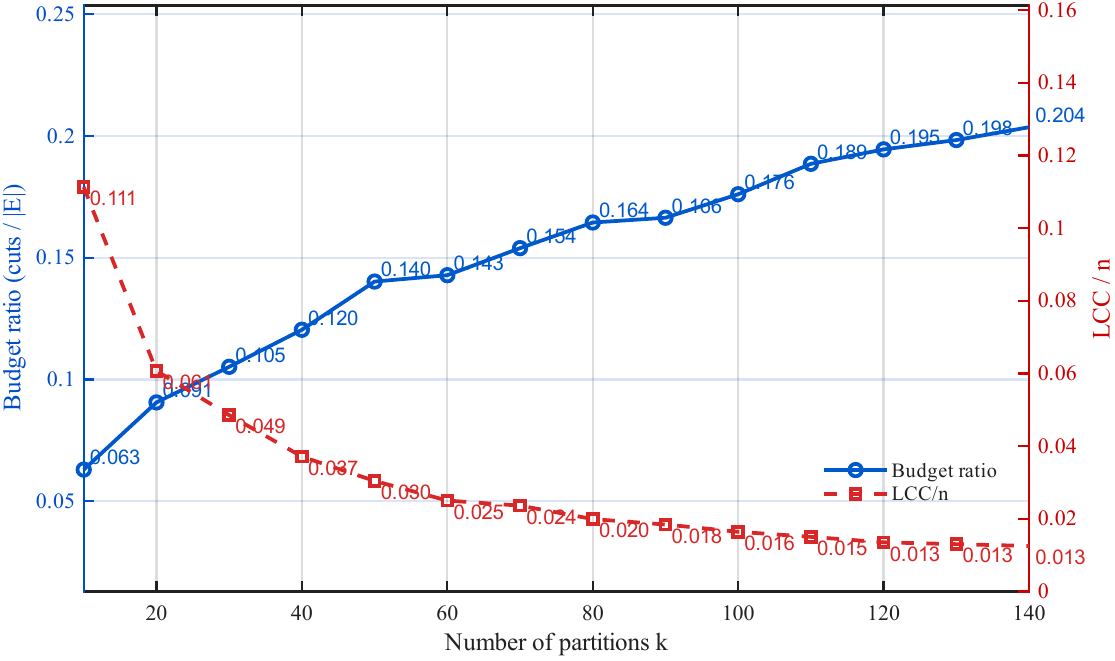}
  \caption{Large-scale dataset results – joint trends of budget ratio and LCC size ratio for different $k$. This result is from a single random planar graph (no standard deviation band), showing a smooth and stable trend.}
  \label{fig:k_budget_lcc_large}
\end{figure}

Figure~\ref{fig:k_budget_lcc_large} presents the budget ratio vs. LCC ratio trend for varying $k$ in the large-scale dataset. This figure is based on a single large planar graph (hence no standard deviation band) and shows a smooth, stable trend: as $k$ increases, $cut/|E|$ increases monotonically while $L_{\max}/|V|$ decreases monotonically. In this case, each $k$-partitioning had an average runtime of around 500 seconds, reflecting the near-linear increase in runtime with graph size (as predicted by the complexity analysis).

\subsection{Complexity and Parallelism Verification}
To verify the complexity analysis, we measured the average runtime per partition $k$ on both the small-scale ($|V|=200$, $|E| \in [250, 550]$) and large-scale ($|V|=10^4$, $|E| \in [1.1\times 10^4, 2.9\times 10^4]$) datasets. The records show that under the large-budget path, the average runtime increases from about 5 s on the small graphs to about 500 s on the large graphs. The runtime grows approximately linearly with $|E|$, consistent with the complexity $T = O(m\,|E|\,|K|)$.

Considering constant factors and parallelization overhead, the empirical wall-clock time can be described by 
\[T_{\text{wall}} \approx \frac{c_1\, m\, |E| + c_2\, m\, |E|\, |K|}{p} + T_{\text{overhead}},\] 
where $p$ is the number of parallel threads/devices, $T_{\text{overhead}}$ includes task partitioning and result aggregation time, $c_1$ corresponds to the small-budget path’s single “sample + partition” runs, and $c_2$ corresponds to the large-budget path’s multiple “sample + partition” runs across $|K|$ values.

\textit{Small vs. large budget dominant costs.} (i) \textbf{Small-budget path ($cut < |F^{(2)}|$):} only runs on a subgraph $G_s$ of size $n_s = \hat{\alpha} n$ (with $|E_s| \approx \hat{\alpha}|E|$), so the dominant cost is $T_{\text{small}} = O(m\,|E_s|) \approx O(m\,\hat{\alpha}\,|E|)$, scaling roughly linearly with $\hat{\alpha}$. (ii) \textbf{Large-budget path ($cut \ge |F^{(2)}|$):} needs to search over $K = \{\hat{k}_0 - \Delta, \dots, \hat{k}_0 + \Delta\}$, with dominant cost $T_{\text{large}} = O(m\,|E|\,|K|)$, which for $\Delta$ of 1 or 2 is only a constant-factor overhead.

\textit{Parallelization and speedup.} Different random seed trials and different $k$ partitions are independent, so the algorithm can be parallelized along both the “seed dimension” and “partition dimension.” Assuming a parallelizable portion $\rho \approx 1$ (in practice $\rho$ is very close to 1), by Amdahl’s law the speedup $S_p = \frac{1}{(1-\rho) + \rho/p}$ and efficiency $E_p = S_p/p$. In large $p$ and fixed $|K|$, $S_p$ grows nearly linearly with $p$, and $E_p$ is mainly affected by $T_{\text{overhead}}$ and I/O. In our experiments, as $p$ increased and $|K|$ was fixed, we observed $S_p$ scaling roughly linearly with $p$, with $E_p$ slightly reduced by overhead.

\textit{Space complexity and scalability.} The algorithm requires $O(|V|+|E|)$ memory to store the graph. The core tree partition step only needs to store linear-size arrays like $parent(\cdot)$, $parent\_edge(\cdot)$, $children(\cdot)$, and $subtree(\cdot)$, adding an extra $O(|V|)$ overhead. Thus, the overall memory complexity is $Mem = O(|V|+|E|)$, which scales well for large planar graphs (sparse, $|E| = O(|V|)$).

\textit{Empirical consistency with graph size.} Comparing the $\sim 5$ s runtime on small graphs vs.\ $\sim 500$ s on large graphs with the ratio of their edge counts, we find an approximately proportional relationship. Considering that higher $k$ in large graphs requires more candidate edge checks and connectivity updates (absorbed into $c_2$), the observed trend still aligns with the $O(m\,|E|\,|K|)$ dominant complexity.

\subsection{Discussion}

\begin{enumerate}[label=(\arabic*), leftmargin=0pt, itemindent=1.5em, labelsep=0.5em, itemsep=1em]

\item \textbf{Monotonic and smooth budget--fragmentation relationship.} 
As the budget increases, $L_{\max}/|V|$ steadily decreases. 
In the small-budget regime, the drop is steeper (higher per-edge efficiency $\eta$). 
When entering the large-budget regime and gradually increasing $k$, diminishing marginal returns become evident: 
further fragmentation requires additional finishing cuts and cross-partition checks, so the improvement in $L_{\max}$ per edge removed weakens.

\item \textbf{Path division and complementarity.} 
The small-budget path, through the density-informed estimate $\hat{\alpha}$, confines computation to an induced subgraph, achieving a rapid LCC reduction with low cost. 
The large-budget path, guided by the near-linear prior $|F^{(k)}| \approx |F^{(2)}| + s(t)(k-2)$, searches globally in a short interval around $\hat{k}_0$ to ensure that when the budget is sufficient, we approach a balanced multi-component fragmentation.

\item \textbf{Robustness and variance.} 
Across random seeds, the small-scale dataset results have relatively narrow standard deviation bands, and the large-scale dataset (single graph) shows a smooth trend. 
Overall, the diversity of spanning trees provides an averaging effect against local structural biases, avoiding chance extremes caused by any single skeleton.

\item \textbf{Runtime and parallelization consistency.} 
Measured times align with the predicted $O(m\,|E|\,|K|)$: 
increasing $m$ (for stability) or increasing $|K|$ (for robustness) linearly increases runtime; 
parallelizing across seeds/partitions yields nearly linear reduction in wall-clock time. 
For practical deployment, given a fixed computational budget, we recommend keeping $|K|$ small (e.g., $\Delta = 1$ or $2$) and adjusting $m$ to balance solution quality and variance.

\item \textbf{Potential limitations and improvements.}
\begin{itemize}[itemsep=2pt, topsep=2pt]
    \item \textit{Model prior bias:} The slope model $s(t)$ and function $g_{\theta}$ are derived from empirical fitting; in extremely sparse or highly clustered graphs, they might underestimate the necessary cuts. Future work could introduce piecewise or fractal density features, or use Bayesian hierarchical regression, to improve generalization to diverse topologies.
    \item \textit{Greedy partition optimality:} In chain-like or star-like skeletons, the post-order greedy cut may produce slight imbalance. Adding a local swap or heuristic backtracking in the adjustment phase could improve overall balance and further reduce $L_{\max}$.
    \item \textit{Cross-partition edge handling:} After cutting the tree, the presence of non-tree edges can affect the final $L_{\max}$. A lightweight second-phase filtering of cross-partition edges (e.g., using approximate edge betweenness) could yield a more balanced fragmentation.
    \item \textit{Broader evaluation metrics:} Currently we focus on $L_{\max}/|V|$ as the main metric. Future work could incorporate metrics like average component size or entropy-based balance to provide a more comprehensive assessment of network dismantling effects.
\end{itemize}

\end{enumerate}

\section{Conclusion and Future Work}

We presented a spanning-tree-skeleton-based dual-path integrated algorithmic framework for planar graph disruption under an edge budget constraint, aiming to balance fragmentation, efficiency, and interpretability. The core idea is to approximate the network’s connectivity backbone with multiple spanning tree samples, execute a one-time balanced cut on the tree to get the baseline cost $|F^{(2)}|$, and then adaptively choose either a “density-informed” small-budget path or a “slope-prior” large-budget path based on the budget, thereby unifying fast local estimation and global balanced optimization.

\textbf{Main conclusions:}
\begin{enumerate}[label=(\roman*), leftmargin=1.5em, itemsep=1em]
\item \textit{Framework effectiveness.} Experimental results, as illustrated in Figures~\ref{fig:smallbudget_demo} and \ref{fig:planar_partition_examples} through \ref{fig:k_budget_lcc_large}, demonstrate that our method steadily reduces the LCC ratio $L_{\max}/|V|$ under different budgets. In the small-budget regime, the algorithm quickly achieves significant fragmentation benefits via subgraph estimation; in the large-budget regime, the slope model $s(t)$ predicts the optimal number of partitions $\hat{k}$, yielding an approximately linear budget--fragmentation relationship. Although direct comparative experiments with baseline algorithms (random edge removal, betweenness attack, and min-cut approximation) have not yet been conducted, theoretical analysis indicates that the proposed framework is expected to exhibit \emph{potential advantages} in fragmentation balance, per-budget efficiency, and computational cost.

\item \textit{Complexity and scalability.} The algorithm’s theoretical complexity is $O(m\,|E|\,|K|)$, with $m$ the number of spanning tree samples and $|K|$ the search width around $k$. Since “random seed trials” and “different $k$ partitions” can be executed independently, the framework naturally supports parallelization in the seed and partition dimensions, with actual speedup close to linear. Measured runtime grows roughly linearly with graph size (edge count), consistent with theory; memory overhead is $O(|V|+|E|)$, easily scaling to planar graphs with over $10^4$ nodes.

\item \textit{Structural interpretability and generality.} By using a spanning tree skeleton to capture the global connectivity backbone, the algorithm provides a clear geometric interpretation for “where to cut” and “why to cut.” Moreover, since the core modules depend only on spanning tree construction and post-order partition logic, the algorithm can be seamlessly transferred to other spatially embedded networks (such as road, power, or communication networks), demonstrating strong practical usability and cross-domain potential.
\end{enumerate}

\textbf{Limitations and future directions:} Despite the high efficiency and stability of our method on planar graphs, several improvements are possible:
\begin{enumerate}[label=(\alph*), leftmargin=1.5em, itemsep=0.8em]
\item The density-informed function $g_{\theta}$ and slope model $s(t)$ rely on prior fitting and may introduce bias in extremely sparse or highly clustered graphs. Future work can explore adaptive regression models (e.g., piecewise polynomials, kernel-based non-parametric fits, or small neural regressors) to enhance generalization.

\item The current post-order greedy partitioning might yield slight imbalance in long-chain or star-like structures; adding a local swap or heuristic backtracking step during the adjustment phase could improve balance and further reduce $L_{\max}$.

\item After tree cuts, non-tree (cross-partition) edges may still connect components; a lightweight cut refinement based on edge centrality could improve final fragmentation quality.

\item We focused on $L_{\max}/|V|$ as the primary metric; incorporating additional metrics (average component size, fragmentation entropy, etc.) in the evaluation could provide a more comprehensive view of network disruption impacts.
\end{enumerate}

In summary, the proposed “spanning tree skeleton + dual-path adaptivity” framework provides a unified and efficient approach for planar graph dismantling and spatial network robustness analysis. This idea is not only practically meaningful for tasks such as network interdiction planning and cascading failure evaluation, but also offers a new computational perspective for structural simplification and hierarchical understanding of complex networks.

\end{document}